\documentclass{pasj00}

\SetRunningHead{Richmond et al.}{Proper motions with Subaru}

\usepackage{times}

\begin{document}

\title{Proper motions with Subaru II. 
           A sample in the Subaru/XMM-Newton Deep Survey field
    \thanks{Based in part on data collected at the Subaru Telescope, 
     which is operated by the National Astronomical Observatory of Japan.}
}
\author{Michael W. \textsc{Richmond}
}
\affil{%
   Physics Department, Rochester Institute of Technology, \\
   Rochester, NY 14623, USA}
\email{mwrsps@rit.edu}
\author{Tomoki \textsc{Morokuma}}
\affil{Research Fellow of the Japan Society for the Promotion of Science, \\
Optical and Infrared Astronomy Division, National Astronomical Observatory of Japan, \\
        2-21-1 Osawa, Mitaka, Tokyo 181-8588, Japan}
\email{tomoki.morokuma@nao.ac.jp}
\author{Mamoru \textsc{Doi}}
\affil{Institute of Astronomy, Graduate School of Science, University of Tokyo, \\
          2-21-1, Osawa, Mitaka, Tokyo 181-0015, Japan}
\email{doi@ioa.s.u-tokyo.ac.jp}
\author{Yutaka \textsc{Komiyama}}
\affil{Optical and Infrared Astronomy Division, National Astronomical Observatory of Japan, \\
        2-21-1 Osawa, Mitaka, Tokyo 181-8588, Japan}
\email{komiyama@subaru.naoj.org}
\author{Naoki \textsc{Yasuda}}
\affil{Institute for the Physics and Mathematics of the Universe, 
        University of Tokyo, \\
        5-1-5 Kashiwa-no-ha, Kashiwa, Chiba 277-8568, Japan}
\email{yasuda@icrr.u-tokyo.ac.jp}
\and
\author{Sadanori \textsc{Okamura}}
\affil{Department of Astronomy and Research Center for the Early Universe,
            School of Science, University of Tokyo \\
        7-3-1 Hongo, Bunkyo, Tokyo 113-0033, Japan}
\email{okamura@astron.s.u-tokyo.ac.jp}

\KeyWords{stars: kinematics ${}$ --- Galaxy: kinematics and dynamics${}$ --- Galaxy: structure${}$}

\maketitle

\begin{abstract}
We search for stars with proper motions in a set of 
deep Subaru images, covering about 0.48 square degrees
to a depth of $i' \simeq 26$, 
taken over a span of five and a half years.
We follow the methods described in 
\citet{Richmond2009} 
to reduce and analyze this dataset.
We present a sample of
69 stars with motions of high significance,
and discuss briefly the populations
from which they are likely drawn.
Based on photometry and motions alone,
we expect that 14 of the candidates
may be white dwarfs.
Our candidate with the largest proper motion
is surprisingly faint and likely to prove
interesting:
its colors and motions suggest that it might be
an M dwarf moving at over 500 km/sec or
an L dwarf in the halo.
\end{abstract}

\section{Introduction}

This paper continues our effort
to use
the Subaru telescope to search for proper motions
among very faint stars.
Our first paper
(\cite{Richmond2009})
examined a series of images of the
Subaru Deep Field (SDF)
(\cite{Kashikawa2004}),
located close to the
Northern Galactic Pole 
($l = 37{\rlap.}^{\circ}6$,
$b = +82{\rlap.}^{\circ}6$).
We report here on a similar analysis of images
of the Subaru/XMM-Newton Deep Survey (SXDS) field
(\cite{Sekiguchi2005},
 \cite{Furusawa2008},
 \cite{Morokuma2008}),
which lies at high galactic latitude
in the southern galactic hemisphere
($l \simeq 170^{\circ}$,
$b \simeq -60^{\circ}$).
As in our earlier work,
we are taking advantage of a dataset 
compiled for the study of very distant
extragalactic objects;
the long time coverage necessary to detect
supernovae and AGN provides us with the baseline
needed to measure proper motions
for a significant number of stars in our own Milky Way.
We pay special attention to members of the faintest
stellar populations, white dwarfs.

Since we will follow very closely the methods 
used in our analysis of the SDF,
we urge the reader to consult
\citet{Richmond2009}
for a detailed description of some procedures
we may mention only briefly here.
However, the SXDS dataset differs from the SDF dataset
in one very important way: 
it consists of a mixture of images taken through 
two passbands, the
Suprime-Cam $R_c$ and $i'$ 
(\cite{Miyazaki2002}),
while the SDF data was all taken through $i'$.
This inhomogeneity complicates efforts to 
determine the selection effects which define
our sample of moving objects;
therefore, we discuss in depth our tests of
the magnitudes and motions to which our analysis
of the SXDS is reasonably complete.
One of our goals in this project is to compare
observed sets of moving stars to those predicted by
various models of stellar populations in the Milky Way.
We will need
a good understanding of the selection 
effects on the observations
in order to compare them fairly to models
in a future paper.

Section 2
describes the observations and the steps 
we took to convert the raw images into 
clean, seamless mosaic images.
We list in section 3 our procedures for
finding and measuring the properties of stars
in the images,
yielding a list of formal motions for tens of thousands
of stars.
Most of these motions, of course, were not
significantly different from zero,
and so we discuss in section 4
our techniques for selecting a small subset
of stars with significant proper motions.
We used artificial stars inserted into our images
to estimate the completeness of our sample
as a function of magnitude and motion,
and, to a limited degree, color.
In section 5, 
we briefly compare our sample of moving stars with 
those found in the SDF and predicted in the 
the Besan\c{c}on model
(\cite{Robin2003}).
We also highlight one very interesting star,
which combines a large proper motion with a very
faint apparent magnitude.

\section{Observations}

\begin{figure}
   \begin{center}
     %\FigureFile(80mm,80mm){sxdsregions.eps}
     \FigureFile(80mm,80mm){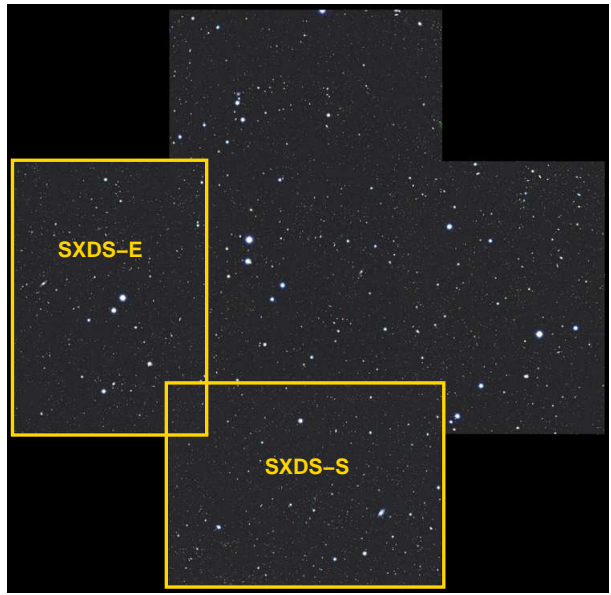}
   \end{center}
   \caption{Subregions within the SXDS in which 
              we measured proper motions.  
              North is up, East to the right.
              The entire field is about 1.3 degrees on a side.
              \label{fig:sxdsregions} }
 \end{figure}

The Subaru/XMM-Newton Deep Survey (SXDS) involves 
sensitive measurements across a wide range of wavelengths over a
region of about 1.3 square degrees
(\cite{Sekiguchi2005},
 \cite{Furusawa2008}).
Since the main goal of the project is to study
extragalactic objects,
the field is located at relatively high galactic latitude
($b = -60^{\circ}$).
We analyze a subset of the optical images
taken with
the Subaru 8.2-meter telescope
and 
Suprime-Cam camera
(\cite{Miyazaki2002}).
The regions 
in the southern (SXDS-S) 
and eastern (SXDS-E) 
sections 
of the survey
(see Figure \ref{fig:sxdsregions}
and Table \ref{table:regions})
were observed 
at least 11 times
over the period 
September, 2002,
to January, 2008.
We provide a list of the observations
in 
Table
\ref{table:journal}.
Note that data taken on
September 29 and September 30, 2002,
were combined to form a single composite image.

 \begin{longtable}{l c c c c}
   \caption{Regions in the SXDS (equinox J2000)}
   \label{table:regions}
   \hline \hline
   Name \quad  & min RA & max RA & min Dec & max Dec \\
   \endfirsthead
   \hline \hline
   \endhead
   \hline 
   \endfoot
   \hline 
   \endlastfoot
 
   \hline 
   SXDS-E         & 02:18:55 & 02:20:34 & -05:17:24 & -04:43:12 \\
   SXDS-S         & 02:16:37 & 02:18:54 & -05:32:22 & -05:06:50 \\

   \hline
   
 \end{longtable}
 
 \begin{longtable}{l c c r}
   \caption{Observations of the SXDS}
   \label{table:journal}
   \hline \hline
   UT Date \quad  & Julian Date - 2,450,000 & 
              Filter & \phantom{} Exptime (seconds) \phantom{} \\
   \endfirsthead
   \hline \hline
   \endhead
   \hline 
   \endfoot
   \hline 
   \endlastfoot
 
   \hline 
   & \phantom{fillingspace} SXDS-E    & & \\
   2002 Sep 29, 30  & 2548.5  &  $i'$  & 3000 \qquad\qquad  \\
   2002 Nov 1       & 2579.9  &  $i'$  & 3000 \qquad\qquad  \\
   2002 Nov 1       & 2580.0  &  $R_c$ & 1500 \qquad\qquad  \\
   2002 Nov 2       & 2580.8  &  $i'$  & 1800 \qquad\qquad  \\
   2002 Nov 9       & 2587.8  &  $i'$  & 2820 \qquad\qquad  \\
   2002 Nov 9       & 2588.0  &  $R_c$ &  210 \qquad\qquad  \\
   2002 Nov 29      & 2608.0  &  $i'$  & 1800 \qquad\qquad  \\
   2003 Sep 22      & 2905.0  &  $i'$  & 6000 \qquad\qquad  \\
   2003 Oct 2       & 2915.1  &  $i'$  & 1271 \qquad\qquad  \\
   2003 Oct 21      & 2934.1  &  $i'$  & 1400 \qquad\qquad  \\
   2005 Sep 28      & 3642.0  &  $i'$  & 3600 \qquad\qquad  \\
   2008 Jan 9       & 4474.8  &  $R_c$ & 2400 \qquad\qquad  \\
     \\

    & \phantom{fillingspace} SXDS-S  & & \\
   2002 Sep 29, 30  & 2548.4  &  $i'$  & 3000 \qquad\qquad  \\
   2002 Nov 1       & 2579.9  &  $i'$  & 3600 \qquad\qquad  \\
   2002 Nov 1       & 2580.0  &  $R_c$ & 2400 \qquad\qquad  \\
   2002 Nov 2       & 2580.8  &  $i'$  & 1800 \qquad\qquad  \\
   2002 Nov 9       & 2587.8  &  $i'$  & 2580 \qquad\qquad  \\
   2002 Nov 9       & 2588.0  &  $R_c$ & 2400 \qquad\qquad  \\
   2002 Nov 29      & 2608.0  &  $i'$  & 1500 \qquad\qquad  \\
   2003 Sep 22      & 2905.0  &  $i'$  & 4500 \qquad\qquad  \\
   2003 Oct 2       & 2915.1  &  $i'$  & 2040 \qquad\qquad  \\
   2005 Sep 28      & 3642.0  &  $i'$  & 2040 \qquad\qquad  \\
   2008 Jan 9       & 4474.7  &  $R_c$ & 2400 \qquad\qquad  \\

 \end{longtable}

We followed the procedure described in
\citet{Richmond2009} 
to reduce the raw images 
and combine them to form a single seamless
mosaic for each night.
The FWHM of the mosaics varied from
$0{\rlap.}^{''}59$ 
on the best night
to 
$1{\rlap.}^{''}11$ 
on the worst,
but since the plate scale was
$0{\rlap.}^{''}202$ per pixel,
all images were adequately sampled.

The limiting magnitude varied slightly,
but on average was about $i' \sim 26.0$
(see Section 3).
These measurements reach about $0.5$ magnitudes
deeper than those reported in 
\citet{Richmond2009};
combined with the larger sky coverage,
this study encompasses a larger volume.
  We estimate the effective volume for
  stars of absolute magnitude
  $M_V = +16.5$ to be 130000 cubic
  parsecs, about four times as large
  as that of 
  \citet{Richmond2009}
  (the value of 14000 cubic parsecs shown
  in Table 1 of that paper was an error;
  the proper value is 28000 cubic parsecs).

\section{Searching for moving objects}

We now describe briefly the steps we took to reduce
the clean mosaic images into a list of starlike objects,
and then to find objects with significant motions.
The reader can find a detailed description of our
methods in 
\citet{Richmond2009}.

Our first step was to identify and measure the
properties of starlike objects.
We used the ``stars'' program in the 
{\it XVista} package
(\cite{Treffers1989})\footnote{http://spiff.rit.edu/tass/xvista}
to find objects with
$0{\rlap.}^{''}6 < {\rm FWHM\ } < 1{\rlap.}^{''}4$.
The images taken on September 22, 2003, 
contained the largest number of objects:
79000 in SXDS-E and 86000 in SXDS-S.
We therefore made this the fiducial epoch
for matching and astrometry.
We broke each image into overlapping subsections roughly
$200^{''}$ on a side and used the
{\it match} package
(\cite{Droege2006})\footnote{http://spiff.rit.edu/match}
to transform the subsections in each 
image to the fiducial's 
coordinate system.
We looked for matches between epochs using a maximum
separation of 
$1{\rlap.}^{''}0$ 
from the fiducial position of each star.
This places an upper limit of about
$0{\rlap.}^{''}23$ per year 
on the proper motions we could detect,
but, as we will show later,
this does not have a strong effect on the results.

% now talk about completeness

\begin{figure}
   \begin{center}
     %\FigureFile(80mm,80mm){completenessapqr.eps}
     \FigureFile(80mm,80mm){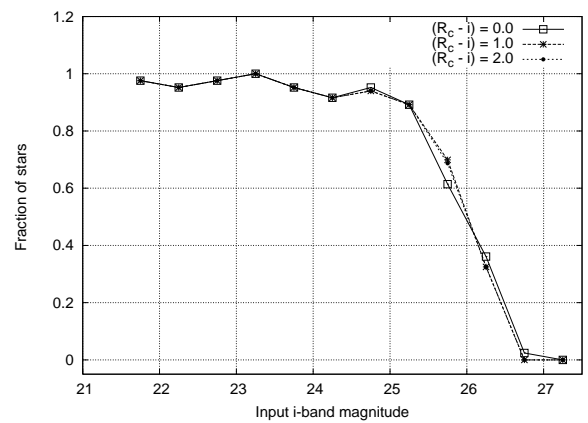}
   \end{center}
   \caption{Fraction of artificial stars added to the images
              which were detected and matched in at least 5 epochs.
              \label{fig:completenesspqr} }
 \end{figure}

In order to test the completeness of our object
detection, we used a set of artificial stars
inserted into the images of SXDS-S. 
Because this field was observed one fewer time than
SXDS-E, the limits we derive from it will likely
be conservative estimates for SXDS-E.
The tests are complicated by the mixture of
passbands in the dataset: eight in $i'$ and
three in $R_c$, including the critical final image
taken in 2008.
We therefore ran three sets of tests,
using artificial stars with colors
$(R_c - i')$ of $0.0$, $1.0$ and $2.0$ magnitudes.
For each color, we created 1000 artificial stars
and inserted them at locations known to be free of real
stars in each image.
We then executed our procedures to find and
match stars.
We required stars to appear in at least five epochs
to qualify for further study.
Figure \ref{fig:completenesspqr}
indicates that the fraction of artificial
stars entering our proper motion study
dropped to 50\% at $i' \simeq 26.0$
regardless of color.

\begin{figure}
   \begin{center}
     %\FigureFile(80mm,80mm){compareposmedianuncert2.eps}
     \FigureFile(80mm,80mm){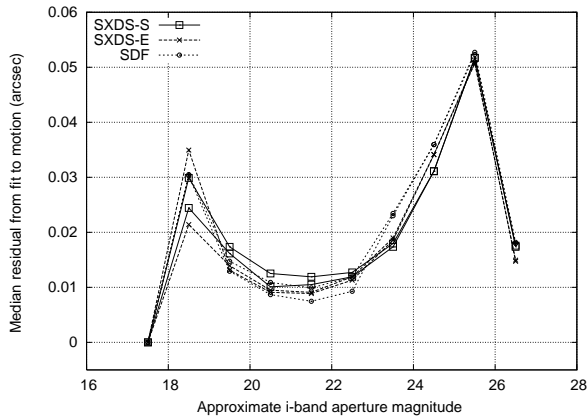}
   \end{center}
   \caption{Comparison of residuals from fitted proper motions
              along rows and along columns in the SDF and SXDS.
              \label{fig:movingresiduals} }
 \end{figure}

% now discuss method of fitting proper motion via straight line
A large number of objects qualified for
the next step by appearing in at least five images:
47589 in SXDS-S and 45500 in SXDS-E.
We considered the row and column positions
of each object separately,
making a linear fit to each as a function of time.
As described in 
\citet{Richmond2009},
we derived the one-dimensional
motion in rows and columns,
as well as the uncertainty in each motion.
% This goes after talk of fitting motions to a straight line
Figure \ref{fig:movingresiduals}
indicates that
the precision of our measurements of
position in the SXDS
is roughly the same as that of our measurements
in the SDF.
The median deviation in each direction
from the fitted motion ranged from
$0{\rlap.}^{''}01$ 
for bright unsaturated stars 
to 
$0{\rlap.}^{''}05$ 
for the faintest objects in our sample.

\section{Choosing objects with significant motions}

% separate stars by significance of motion
We defined statistics $S_{\rm row}$
and $S_{\rm col}$ as the ratios of motion
to uncertainty in each direction,
and computed the overall significance
of each object's motion as
$S_{\rm tot} \equiv \sqrt{ S_{\rm row}^2 + S_{\rm col}^2 }$.
We chose a conservative threshold,
$S_{\rm tot} \geq 5$,
to create samples of
``candidates for stars with proper motions.''
Only a small number of objects
entered these samples:
40 in the SXDS-S and 33 in the SXDS-E.
The fraction of all objects having
significant motions was only about
$0.7\%$, smaller than the 
$1.4\%$ found in the SDF.
Since the uncertainties in individual
measurements were roughly the same in
both areas, but stars were measured
almost twice as frequently in the SDF,
a fixed value of $S_{\rm tot}$ 
corresponds to a smaller absolute motion
in the SDF than in the SXDS fields.

\begin{figure*}
   \begin{center}
     %\FigureFile(150mm,240mm){artmotionsd.eps}
     \FigureFile(150mm,240mm){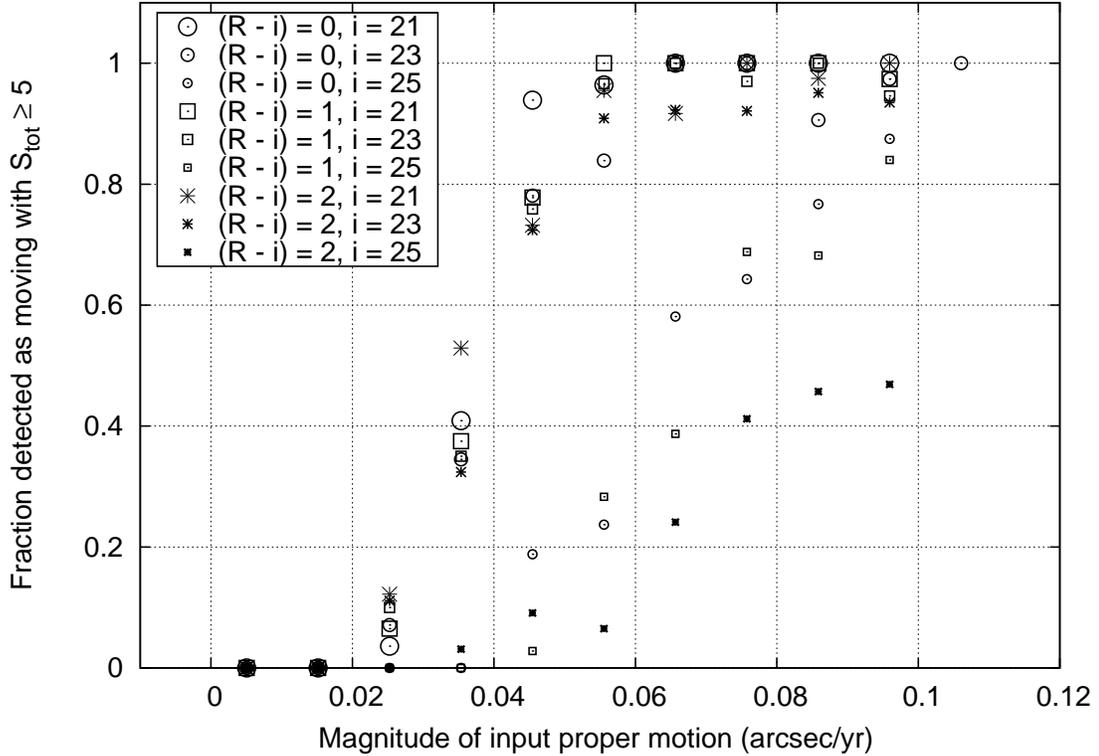}
   \end{center}
   \caption{Results of tests using moving artificial stars to
              check completeness as a function of color and motion.
              \label{fig:pmcolor} }
 \end{figure*}

The time sampling of our survey was far from uniform,
and there was a three-year gap between the penultimate
and final measurement.
Moreover, that final measurement was made through the
$R_c$ filter, rather than the $i'$ filter used for
most of the images.
Since that final image influences the proper motions
strongly, it is possible that there might be some 
color dependency in our proper motion sample.
Therefore, we ran a set of tests using three 
sets of artificial stars:
one group had colors $(R_c - i') = 0$,
one ($R_c - i') = 1$, and one $(R_c - i') = 2$.
We generated 1000 stars in each group with
$i'$-band magnitudes $21 \leq i' \leq 25$
and proper motions 
$0{\rlap.}^{''}0 \leq \mu \leq 0{\rlap.}^{''}10$ 
per year
in random directions.
We inserted these artificial stars into our images,
analyzed them as described above,
and counted the number which appeared in our
output set of stars detected as moving with
significance 
$S_{\rm tot} \geq 5$.
As
Figure \ref{fig:pmcolor}
shows,
the efficiency with which we detect moving 
stars does depend on color for the faintest stars:
we recover faint blue stars at a higher rate than
faint red stars.
At the bright end, our efficiency drops to 
50\% for motions less than about 
$0{\rlap.}^{''}04$ per year for stars of all colors.
Our search in the SXDS is slightly less sensitive 
to stars with small motions than our search in the
SDF, due to the smaller number of images over a
similar span of time.

\begin{figure}
   \begin{center}
     %\FigureFile(80mm,80mm){pmhistb.eps}
     \FigureFile(80mm,80mm){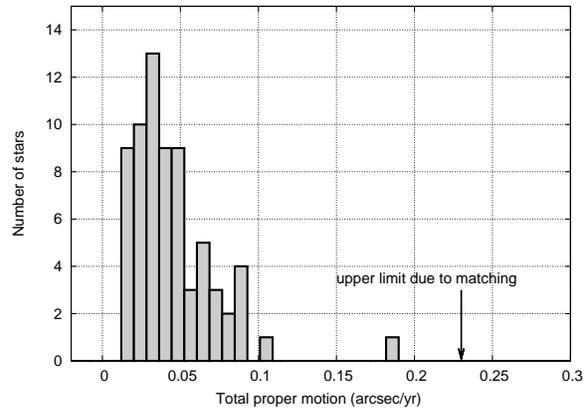}
   \end{center}
   \caption{Distribution of detected proper motions.
              \label{fig:pmhist} }
 \end{figure}

We show in
Figure \ref{fig:pmhist}
the distribution of measured proper motions.
The upper limit of 
$0{\rlap.}^{''}23$ per year 
is set by 
the matching limit of 
$1{\rlap.}^{''}0$
and
the time difference between
our fiducial epoch (Sept 22, 2003)
and the final epoch (Jan 9, 2008).
Since all but a single object
have motions less than half this value,
we conclude that the limit set by our
matching procedures does not bias our
final results.
We will discuss the one object with
$\mu \sim 0{\rlap.}^{''}18$ per year 
in the next section.

% discard stars with bogus motions after inspection by eye
We examined each candidate visually at 
several epochs to verify that the motions
were real.
We found 1 object in the SXDS-S and 3 in the SXDS-E
which had bogus motions due to blends of two
stars or a star and a galaxy.
Discarding those left 39 candidates in the SXDS-S
and 30 candidates in the SXDS-E.

% photometric measurements and re-measurements for a few
%    stars in each field with suspect B,V mags in SXDS catalogs
We extracted corrected isophotal magnitudes
in $B$, $V$, $R_c$ and $i'$ passbands
for each candidate from the catalogs of
\cite{Furusawa2008}
and compared them to the visual appearance
of the candidate in images taken from
the SXDS Data Release 1\footnote{http://step.mtk.nao.ac.jp/sxds/}.
In several cases (5 in SXDS-S, 4 in SXDS-E),
we noticed an obvious discrepancy, always in the same sense:
the catalog magnitude indicated a much brighter object
than actually appeared in the images.
We suspect that the matching procedure used
to create the catalogs may have confused 
neighboring objects with very different colors.
Therefore, we corrected the magnitudes of these
candidates as follows: we measured instrumental magnitudes of 
all candidates in the $B$, $V$, $R_c$ and $i'$ images
of the SXDS Data Release 1 using a small aperture,
$1$ arcsecond in radius.
We examined the difference 
between our instrumental magnitudes and the catalog
magnitudes in each passband;
in all cases, we found a nearly constant offset
for both bright and faint stars, except for a few
outliers.
We then applied this offset
to the instrumental magnitudes of the 
outliers.
The discrepancies were largest in the $B$ and $V$
passbands and almost zero in $i'$.

There is a small region of overlap between the
SXDS-S and SXDS-E, as one can see in 
Figure \ref{fig:sxdsregions}.
We noticed that two of the candidates
in this region were actually the same object,
detected and measured independently in each 
set of images.
This object was one of the photometric outliers
and serves as a test of our ability to compute
accurate magnitudes.
The difference in corrected magnitudes was
$0.03$, $0.04$ and $0.08$ mag in $V$, $R_c$ and $i'$,
respectively.
The annual proper motions derived in each set of images
agreed very well:
$0{\rlap.}^{''}049 \pm 0{\rlap.}^{''}006$ 
versus
$0{\rlap.}^{''}052 \pm 0{\rlap.}^{''}006$ 
in RA,
and
$-0{\rlap.}^{''}011 \pm 0{\rlap.}^{''}004$ 
versus
$-0{\rlap.}^{''}008 \pm 0{\rlap.}^{''}008$ 
in Dec.
However, when we compared the positions for this
object drawn separately from the 
SXDS-E and SXDS-S
catalogs
of 
\citet{Furusawa2008},
we found differences of
$0{\rlap.}^{''}50$ 
in RA and
$0{\rlap.}^{''}25$ 
in Dec.
This may indicate the accuracy of the
positions of the objects shown in
Table \ref{table:sample}.

\begin{figure}
   \begin{center}
     %\FigureFile(120mm,120mm){pmsignifeboth.eps}
     \FigureFile(120mm,120mm){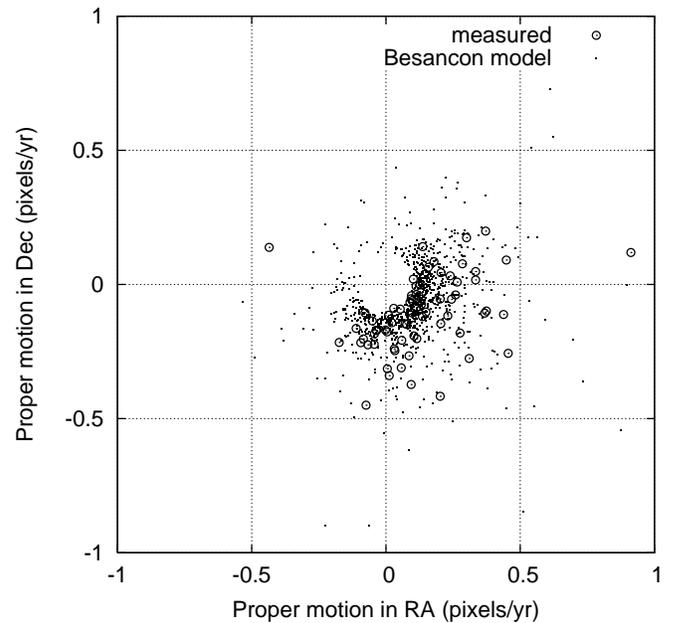}
   \end{center}
   \caption{Comparison of measured and model proper motions
              for stars in the SXDS.
              \label{fig:pmsignif} }
 \end{figure}

% final result: sample of stars with significant motion and good mags
%    put figure showing direction of motions here 
The result is a sample of 68 stars with 
significant proper motions and good photometry
in four passbands.
We list these stars in
Table \ref{table:sample}.
As a final check on the reality of the
motions, we can compare the measured motions
to those predicted by the
Besan\c{c}on model 
of stellar populations
within the Milky Way Galaxy.
Figure \ref{fig:pmsignif}
shows that our measured motions
are predominantly in 
the southeastern direction,
agreeing with the model.

 \begin{longtable}{l l l r r r r r r }
   \caption{First sample of proper motion candidates in the SXDS}
   \label{table:sample}
   \hline \hline
   ID & RA\footnotemark[$*$] & Dec\footnotemark[$*$] & 
         $V$\footnotemark[$**$] $\thinspace $ & 
         $R_c$\footnotemark[$**$] $\thinspace $ & 
         $i'$\footnotemark[$**$] $\thinspace $ & 
         $z'$\footnotemark[$**$] $\thinspace $ & 
         RA PM\footnotemark[$\dagger$] \qquad &  
         Dec PM\footnotemark[$\dagger$] \qquad  \\ 
   \endfirsthead
   \hline \hline
   \endhead
   \hline 
   \endfoot
   \hline 
   \endlastfoot
 
   \hline 
 
 SXDSPM J021651.8-053629 &     34.21609 &     -5.60829 & 20.099 & 19.287 & 18.322 & 17.401 &   0.048 $\pm$  0.006 &  0.006 $\pm$  0.006 \\
 SXDSPM J021658.4-051453 &     34.24373 &     -5.24807 & 20.309 & 20.143 & 20.132 & 20.066 &  -0.009 $\pm$  0.008 & -0.037 $\pm$  0.006 \\
 SXDSPM J021702.0-052310 &     34.25844 &     -5.38629 & 21.361 & 20.552 & 19.495 & 18.638 &  -0.014 $\pm$  0.004 & -0.045 $\pm$  0.006 \\
 SXDSPM J021705.9-053140 &     34.27470 &     -5.52778 & 20.960 & 20.232 & 19.390 & 18.657 &   0.054 $\pm$  0.006 &  0.002 $\pm$  0.002 \\
 SXDSPM J021706.2-052737 &     34.27605 &     -5.46043 & 22.959 & 22.283 & 21.704 & 21.334 &   0.026 $\pm$  0.002 & -0.000 $\pm$  0.002 \\
 SXDSPM J021706.6-052724 &     34.27770 &     -5.45692 & 19.473 & 19.491 & 19.593 & 19.467 &   0.090 $\pm$  0.002 &  0.018 $\pm$  0.006 \\
 SXDSPM J021707.2-052427 &     34.28022 &     -5.40771 & 22.629 & 21.713 & 19.924 & 18.797 &  -0.010 $\pm$  0.006 & -0.027 $\pm$  0.002 \\
 SXDSPM J021708.0-051419 &     34.28370 &     -5.23883 & 21.888 & 21.224 & 20.844 & 20.512 &   0.025 $\pm$  0.002 & -0.009 $\pm$  0.004 \\
 SXDSPM J021708.6-052501 &     34.28613 &     -5.41700 & 21.210 & 20.640 & 20.340 & 20.056 &   0.021 $\pm$  0.002 & -0.039 $\pm$  0.006 \\
 SXDSPM J021709.9-052437 &     34.29164 &     -5.41042 & 19.359 & 18.949 & 18.765 & 18.121 &   0.006 $\pm$  0.010 & -0.048 $\pm$  0.004 \\
 SXDSPM J021712.5-052133 &     34.30238 &     -5.35927 & 25.035 & 23.881 & 22.015 & 21.002 &  -0.016 $\pm$  0.004 & -0.021 $\pm$  0.002 \\
 SXDSPM J021719.9-052038 &     34.33308 &     -5.34394 & 23.456 & 22.977 & 21.860 & 21.242 &   0.018 $\pm$  0.002 & -0.014 $\pm$  0.002 \\
 SXDSPM J021720.9-053642 &     34.33718 &     -5.61191 & 22.624 & 22.024 & 21.111 & 20.589 &   0.038 $\pm$  0.004 & -0.012 $\pm$  0.002 \\
 SXDSPM J021722.4-053136 &     34.34335 &     -5.52691 & 25.028 & 23.684 & 21.769 & 20.732 &   0.036 $\pm$  0.006 &  0.017 $\pm$  0.004 \\
 SXDSPM J021730.6-052459 &     34.37758 &     -5.41639 & 19.645 & 19.246 & 19.093 & 18.484 &   0.041 $\pm$  0.006 & -0.029 $\pm$  0.012 \\
 SXDSPM J021732.2-053558 &     34.38418 &     -5.59953 & 20.670 & 20.113 & 19.875 & 19.462 &   0.001 $\pm$  0.006 & -0.036 $\pm$  0.004 \\
 SXDSPM J021734.8-051850 &     34.39506 &     -5.31412 & 21.303 & 20.604 & 19.945 & 19.372 &  -0.007 $\pm$  0.004 & -0.035 $\pm$  0.004 \\
 SXDSPM J021736.1-053647 &     34.40074 &     -5.61307 & 22.646 & 22.264 & 22.066 & 21.865 &   0.058 $\pm$  0.004 &  0.016 $\pm$  0.006 \\
 SXDSPM J021744.6-052116 &     34.43611 &     -5.35462 & 20.260 & 19.765 & 19.535 & 19.017 &   0.027 $\pm$  0.002 & -0.011 $\pm$  0.006 \\
 SXDSPM J021745.3-053035 &     34.43904 &     -5.50984 & 21.983 & 21.348 & 20.487 & 19.975 &   0.020 $\pm$  0.002 & -0.012 $\pm$  0.002 \\
 SXDSPM J021756.5-053530 &     34.48554 &     -5.59187 & 22.863 & 22.261 & 21.071 & 20.434 &   0.092 $\pm$  0.002 & -0.052 $\pm$  0.002 \\
 SXDSPM J021758.2-052304 &     34.49284 &     -5.38466 & 24.030 & 23.477 & 22.989 & 22.585 &   0.019 $\pm$  0.002 & -0.008 $\pm$  0.004 \\
 SXDSPM J021805.6-053254 &     34.52353 &     -5.54855 & 20.584 & 19.971 & 19.668 & 19.177 &   0.046 $\pm$  0.004 & -0.024 $\pm$  0.006 \\
 SXDSPM J021806.6-053612 &     34.52764 &     -5.60350 & 22.747 & 22.552 & 22.428 & 22.232 &  -0.017 $\pm$  0.004 & -0.041 $\pm$  0.006 \\
 SXDSPM J021806.8-053613 &     34.52869 &     -5.60379 & 22.519 & 21.738 & 20.187 & 19.262 &  -0.019 $\pm$  0.006 & -0.044 $\pm$  0.004 \\
 SXDSPM J021809.7-052547 &     34.54078 &     -5.42992 & 24.105 & 23.070 & 21.488 & 20.653 &   0.032 $\pm$  0.002 &  0.013 $\pm$  0.008 \\
 SXDSPM J021818.2-052739 &     34.57622 &     -5.46110 & 20.333 & 19.746 & 19.515 & 19.002 &   0.007 $\pm$  0.004 & -0.050 $\pm$  0.006 \\
 SXDSPM J021819.0-052129 &     34.57918 &     -5.35825 & 21.402 & 20.613 & 19.516 & 18.610 &   0.027 $\pm$  0.004 & -0.024 $\pm$  0.004 \\
 SXDSPM J021830.8-051918 &     34.62859 &     -5.32185 & 23.006 & 22.375 & 21.998 & 21.704 &   0.002 $\pm$  0.002 & -0.023 $\pm$  0.002 \\
 SXDSPM J021833.5-052918 &     34.63963 &     -5.48858 & 22.841 & 21.973 & 20.409 & 19.523 &   0.056 $\pm$  0.004 & -0.037 $\pm$  0.004 \\
 SXDSPM J021835.5-051734 &     34.64794 &     -5.29294 & 22.605 & 21.889 & 21.000 & 20.489 &   0.001 $\pm$  0.002 & -0.063 $\pm$  0.004 \\
 SXDSPM J021840.0-053623 &     34.66670 &     -5.60661 & $ \geq$ 27.3 & 26.085 & 24.381 & 22.618 &   0.184 $\pm$  0.024 &  0.024 $\pm$  0.028 \\
 SXDSPM J021850.4-053200 &     34.71003 &     -5.53342 & 20.818 & 20.126 & 19.467 & 18.740 &   0.023 $\pm$  0.010 & -0.041 $\pm$  0.004 \\
 SXDSPM J021856.8-044522 &     34.73684 &    -4.75612 & 22.112 & 21.248 & 19.816 & 19.002 &   0.020 $\pm$  0.002 & -0.025 $\pm$  0.008 \\
 SXDSPM J021858.4-052439 &     34.74341 &     -5.41092 & 21.530 & 20.574 & 19.157 & 18.121 &   0.075 $\pm$  0.004 &  0.040 $\pm$  0.018 \\
 SXDSPM J021901.2-051403 &     34.75517 &     -5.23426 & 24.560 & 23.221 & 21.297 & 20.255 &   0.049 $\pm$  0.006 & -0.011 $\pm$  0.004 \\
 SXDSPM J021902.7-053628 &     34.76158 &     -5.60784 & 20.670 & 19.851 & 18.860 & 18.082 &  -0.035 $\pm$  0.012 & -0.044 $\pm$  0.004 \\
 SXDSPM J021903.0-053131 &     34.76257 &     -5.52552 & 20.963 & 20.194 & 19.293 & 18.482 &  -0.022 $\pm$  0.004 & -0.033 $\pm$  0.002 \\
 SXDSPM J021903.7-052540 &     34.76557 &     -5.42783 & 22.867 & 22.167 & 20.903 & 20.143 &   0.011 $\pm$  0.004 & -0.019 $\pm$  0.002 \\
 SXDSPM J021908.5-053002 &     34.78549 &     -5.50076 & 21.144 & 20.406 & 19.500 & 18.737 &   0.063 $\pm$  0.002 & -0.056 $\pm$  0.006 \\
 SXDSPM J021911.0-045902 &     34.79612 &    -4.98411 & 21.049 & 21.244 & 21.398 & 21.520 &  -0.015 $\pm$  0.006 & -0.091 $\pm$  0.002 \\
 SXDSPM J021914.8-045312 &     34.81205 &     -4.88669 & 22.594 & 21.743 & 20.394 & 19.608 &   0.023 $\pm$  0.002 & -0.001 $\pm$  0.002 \\
 SXDSPM J021914.8-050120 &     34.81208 &     -5.02240 & 23.248 & 22.601 & 21.223 & 20.488 &   0.027 $\pm$  0.002 & -0.015 $\pm$  0.002 \\
 SXDSPM J021918.6-045833 &     34.82772 &     -4.97598 & 23.953 & 23.643 & 23.461 & 23.240 &   0.041 $\pm$  0.004 &  0.009 $\pm$  0.010 \\
 SXDSPM J021925.1-051610 &     34.85480 &     -5.26966 & 24.550 & 23.320 & 21.791 & 20.888 &   0.023 $\pm$  0.002 & -0.008 $\pm$  0.008 \\
 SXDSPM J021925.7-050051 &     34.85742 &     -5.01442 & 19.927 & 19.363 & 19.074 & 18.511 &   0.019 $\pm$  0.008 & -0.075 $\pm$  0.010 \\
 SXDSPM J021929.2-045819 &     34.87201 &     -4.97220 & 23.146 & 22.281 & 20.512 & 19.556 &   0.061 $\pm$  0.006 &  0.035 $\pm$  0.006 \\
 SXDSPM J021935.5-050422 &     34.89807 &     -5.07289 & 22.303 & 21.430 & 20.061 & 19.267 &  -0.007 $\pm$  0.004 & -0.035 $\pm$  0.004 \\
 SXDSPM J021936.0-051658 &     34.90005 &     -5.28305 & 20.373 & 19.878 & 19.695 & 19.378 &   0.006 $\pm$  0.004 & -0.018 $\pm$  0.002 \\
 SXDSPM J021936.2-050341 &     34.90105 &     -5.06145 & 20.809 & 20.271 & 19.828 & 19.473 &   0.067 $\pm$  0.008 &  0.010 $\pm$  0.006 \\
 SXDSPM J021938.6-045727 &     34.91113 &     -4.95770 & 25.630 & 23.841 & 21.845 & 20.383 &   0.076 $\pm$  0.010 & -0.020 $\pm$  0.004 \\
 SXDSPM J021939.9-045216 &     34.91662 &     -4.87112 & 22.535 & 21.806 & 20.811 & 20.262 &   0.023 $\pm$  0.002 & -0.018 $\pm$  0.004 \\
 SXDSPM J021942.3-044706 &     34.92665 &     -4.78525 & 23.566 & 22.894 & 22.229 & 21.844 &  -0.009 $\pm$  0.008 & -0.045 $\pm$  0.004 \\
 SXDSPM J021943.9-044432 &     34.93307 &     -4.74238 & 22.059 & 21.078 & 19.541 & 18.541 &   0.017 $\pm$  0.002 & -0.054 $\pm$  0.008 \\
 SXDSPM J021952.2-045056 &     34.96781 &     -4.84908 & 21.380 & 20.357 & 19.436 & 18.679 &   0.012 $\pm$  0.002 & -0.042 $\pm$  0.006 \\
 SXDSPM J021953.9-051649 &     34.97485 &     -5.28045 & 22.290 & 21.608 & 20.948 & 20.537 &   0.074 $\pm$  0.004 & -0.022 $\pm$  0.008 \\
 SXDSPM J021958.4-044516 &     34.99338 &     -4.75447 & 20.862 & 20.156 & 19.558 & 19.020 &   0.030 $\pm$  0.004 & -0.021 $\pm$  0.002 \\
 SXDSPM J022001.8-045640 &     35.00766 &     -4.94449 & 21.073 & 20.326 & 19.478 & 18.785 &   0.014 $\pm$  0.008 & -0.029 $\pm$  0.002 \\
 SXDSPM J022005.0-045022 &     35.02086 &     -4.83953 & 20.564 & 19.919 & 19.653 & 19.230 &   0.002 $\pm$  0.002 & -0.069 $\pm$  0.004 \\
 SXDSPM J022008.8-050351 &     35.03694 &     -5.06436 & 22.370 & 21.614 & 20.343 & 19.619 &   0.041 $\pm$  0.006 & -0.011 $\pm$  0.002 \\
 SXDSPM J022009.5-050912 &     35.03992 &     -5.15360 & 22.600 & 21.935 & 21.072 & 20.589 &   0.004 $\pm$  0.004 & -0.029 $\pm$  0.004 \\
 SXDSPM J022010.3-050259 &     35.04295 &     -5.04978 & 25.174 & 24.962 & 24.666 & 24.316 &   0.088 $\pm$  0.014 & -0.023 $\pm$  0.018 \\
 SXDSPM J022013.1-045912 &     35.05474 &     -4.98686 & 21.256 & 20.525 & 19.642 & 18.973 &   0.020 $\pm$  0.002 & -0.021 $\pm$  0.004 \\
 SXDSPM J022013.7-051353 &     35.05711 &     -5.23166 & 24.690 & 23.278 & 21.038 & 19.823 &   0.067 $\pm$  0.006 &  0.003 $\pm$  0.006 \\
 SXDSPM J022017.4-051313 &     35.07290 &     -5.22034 & 23.826 & 23.254 & 22.519 & 22.072 &  -0.000 $\pm$  0.002 & -0.034 $\pm$  0.004 \\
 SXDSPM J022024.0-050606 &     35.10031 &     -5.10171 & 22.865 & 22.476 & 22.198 & 21.979 &   0.041 $\pm$  0.010 & -0.084 $\pm$  0.004 \\
 SXDSPM J022025.7-051028 &     35.10718 &     -5.17455 & 22.754 & 21.994 & 20.646 & 19.921 &   0.016 $\pm$  0.004 & -0.030 $\pm$  0.002 \\
 SXDSPM J022032.5-050834 &     35.13545 &     -5.14289 & 22.599 & 21.864 & 20.859 & 20.311 &   0.021 $\pm$  0.002 &  0.004 $\pm$  0.004 \\

 \hline
 \multicolumn{8}{@{}l@{}}{\hbox to 0pt{\parbox{180mm}{\footnotesize
 \par\noindent
 \footnotemark[$*$] Equinox J2000, epoch 2003.67.
 \par\noindent
 \footnotemark[$**$] Corrected isophotal magnitudes from catalogs of
                      \citet{Furusawa2008}.
 \par\noindent
 \footnotemark[$\dagger$] Proper motions in arcseconds per year.
 }\hss}}

 \end{longtable}

\section{Discussion}

\begin{figure*}
   \begin{center}
     %\FigureFile(150mm,240mm){reducedpmc.eps}
     \FigureFile(150mm,240mm){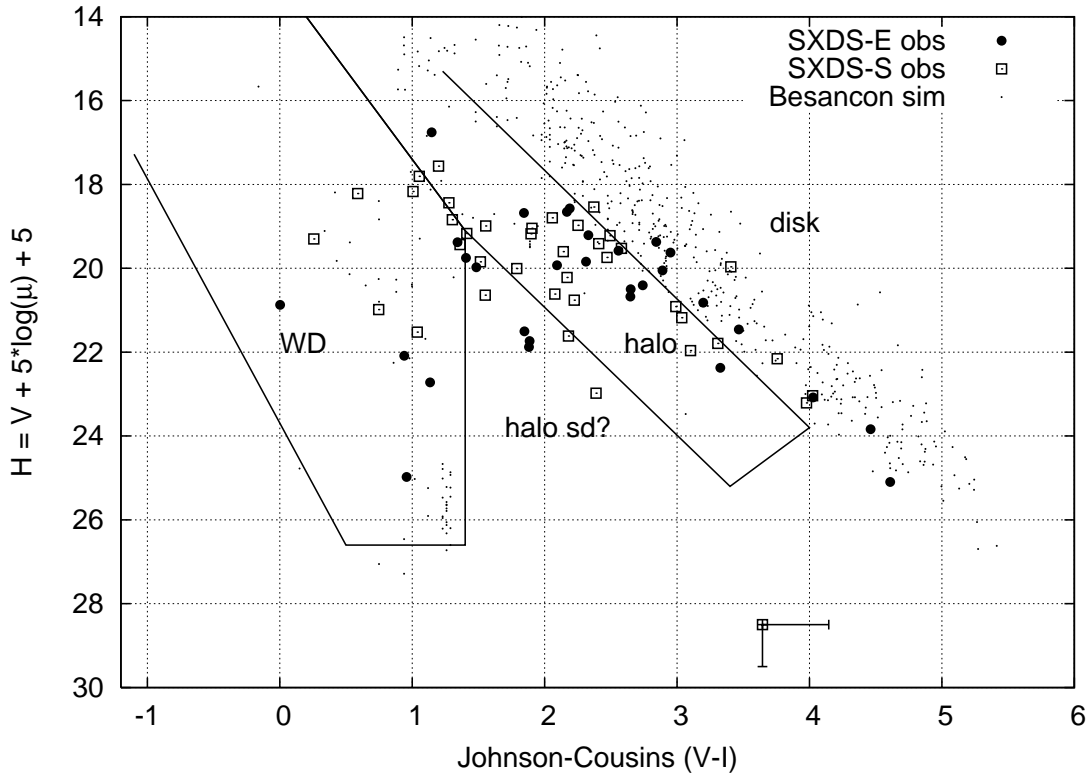}
   \end{center}
   \caption{Reduced proper motion diagram for observed objects
              in the SXDS and simulated objects based on the Besan\c{c}on 
              model.  Real objects have errorbars in both directions,
              though some are too small to see.   
              The object at lower right is placed according to lower limits
              to its $V$-band magnitude; see text for details.
              \label{fig:reducedpm} }
 \end{figure*}

Following 
\citet{Richmond2009},
we computed the reduced proper motion for
the candidates in our sample
in order to separate stars belonging to
different populations.
Figure \ref{fig:reducedpm}
compares stars in our sample
to stars in a simulation of the region
based on the 
Besan\c{c}on model.
The regions drawn in the diagram are
exactly the same as those shown in
Figures 7 and 8 
of 
\citet{Richmond2009}.
We adopt their relationship
\begin{equation}
(V - I) = 0.391 + 1.1145(V_s - i')
\end{equation}
to compare our results to the models.

Of the 68 objects in our sample,
13 lie within the ``white dwarf'' (WD) region
of the diagram,
32 within the ``halo'' region,
12 within the ``disk'' region,
10 within the ``halo sd?'' region,
and 1 falls far from all the rest
(we discuss that outlier below).
The fraction of stars in each region
agree reasonably well with the fractions
found in our survey of the SDF.
Although the fraction of objects in the WD
region (19\%) is somewhat higher than 
in the SDF (9\%), we do not consider the
difference significant, given the small number
of objects in each set, and the number of 
objects lying very close to the 
somewhat arbitrary boundaries of the
WD region in each sample.

We performed multiple simulations using the
Besan\c{c}on model to generate stars 
in the SXDS over
an effective area of two square degrees.
We applied the selection criteria derived in 
Section 3 to these simulated catalogs,
then scaled the results to the actual area
we observed (0.48 square degrees).
We found a total of 12.5 WD in the simulation,
which agrees well with the 13 candidates in 
the WD region of our reduced proper motion diagram.
Over half (56\%) of the WDs in the simulation were members
of the halo,
with most of the remainder (34\%) drawn from the thick disk.

\begin{figure}
   \begin{center}
     %\FigureFile(80mm,80mm){cand5334pos.eps}
     \FigureFile(80mm,80mm){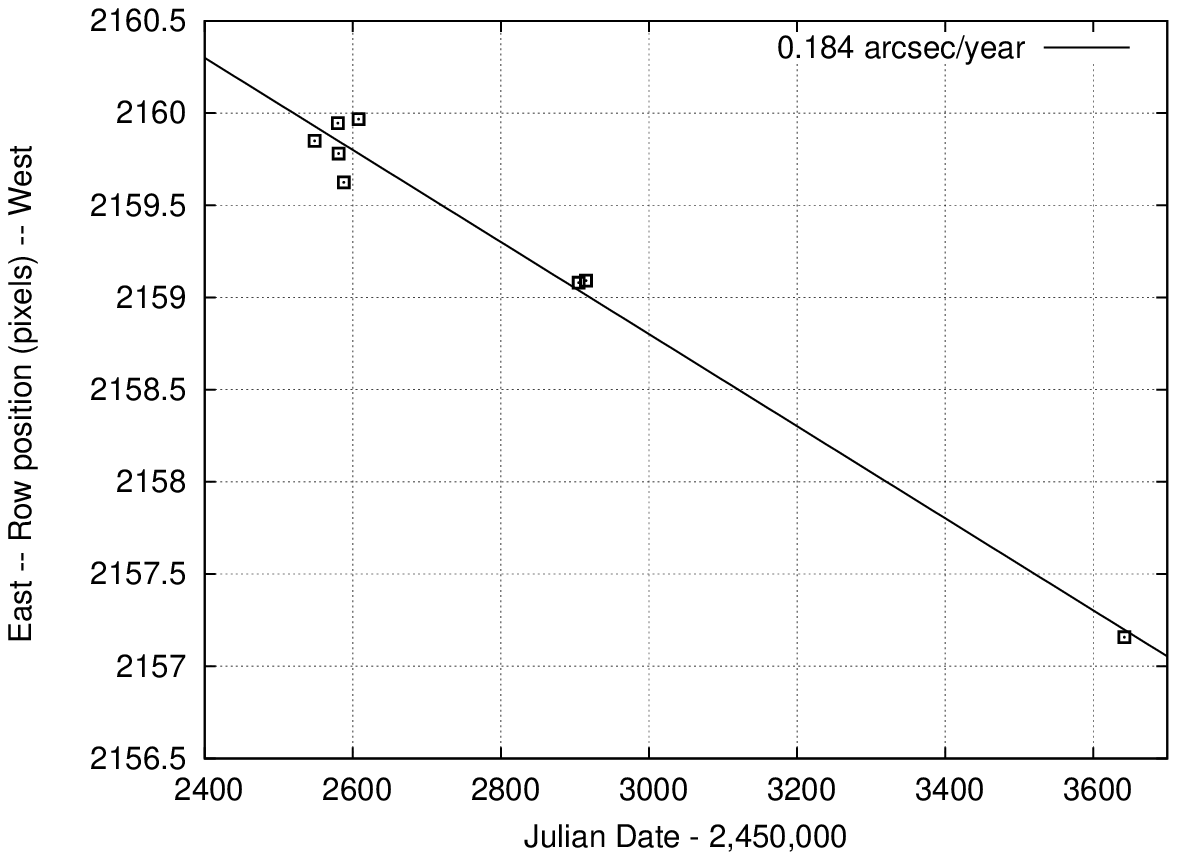}
   \end{center}
   \caption{Motion of SXDSPM J021840.0-053623 
              in Right Ascension over the period
              Sep 29, 2002 to Sep 28, 2005.
              We could not determine a precise position
              of the object in the $R_c$-band image 
              of Jan 9, 2008.
              \label{fig:candmotiondiag} }
 \end{figure}

\begin{figure*}
   \begin{center}
     %\FigureFile(150mm,240mm){candpanels.eps}
     \FigureFile(150mm,240mm){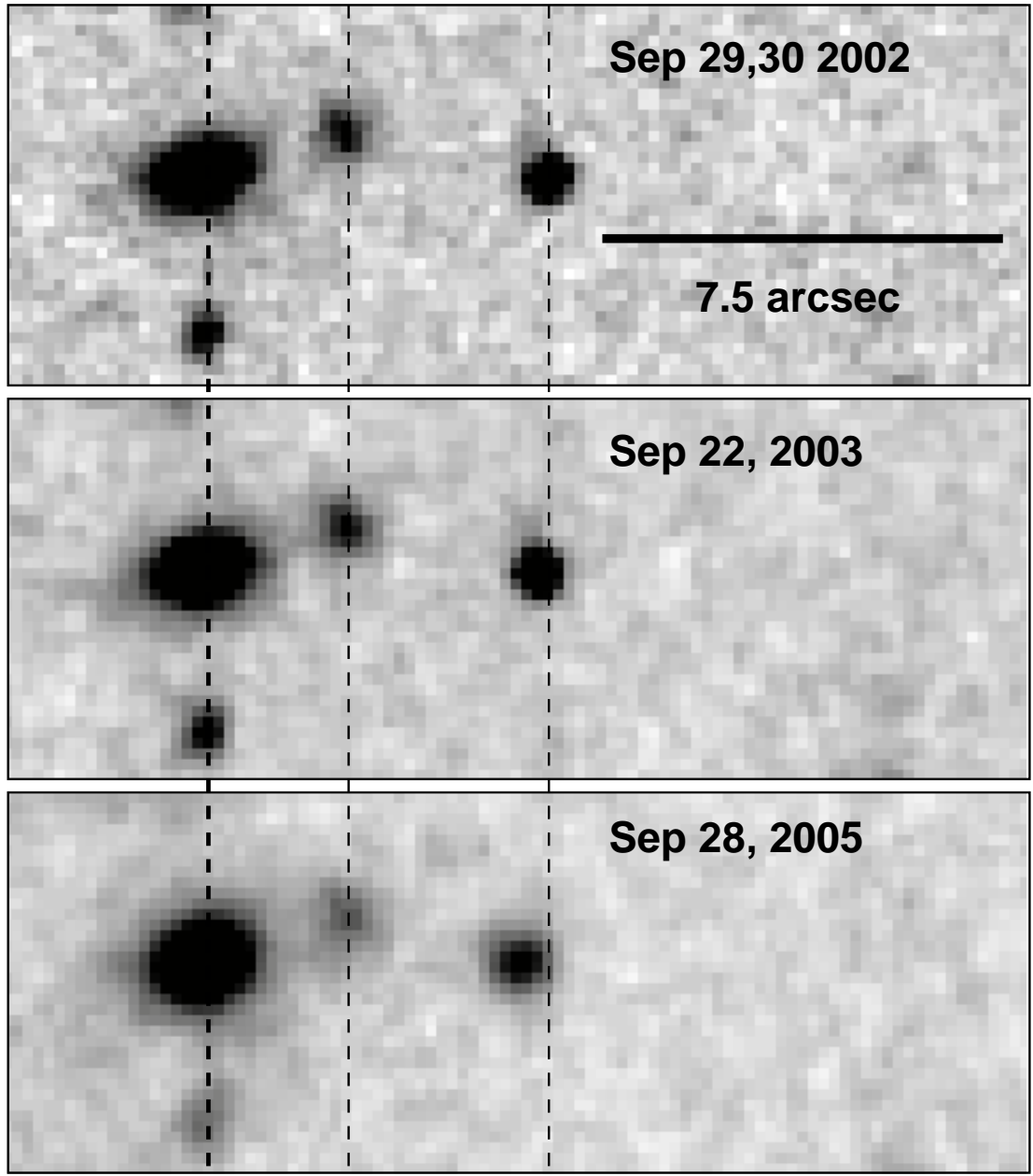}
   \end{center}
   \caption{Images of SXDSPM J021840.0-053623 (at center)
              in $i'$-band over a three-year period.
              North is up and East to the left.
              Vertical lines have been drawn through the
              centers of objects in the 2002 image
              to show relative east-west motion clearly.
              \label{fig:candpanels} }
 \end{figure*}

One of the candidates,
SXDSPM J021840.0-053623,
deserves individual attention.
It has the highest proper motion in our sample,
$0{\rlap.}^{''}186$ per year
(see 
Figures \ref{fig:candmotiondiag}
and
\ref{fig:candpanels}),
and yet is very faint:
$R_c = 26.09$, $i' = 24.38$, $z = 22.62$.
The object is very red: 
our attempt to measure a magnitude 
at the object's position in the 
$V_s$-band image yielded a formal value of 
$V_s \sim 28.3$, but we choose instead 
to quote a lower limit of 
$V_s \geq 27.3$ based on secure
measurements of faint stars nearby.
Its colors $(R_c - i') = 1.69$
and $(i' - z) = 1.76$
place it far from the locus of cool
main sequence stars.
If we use the more reliable $(i' - z)$ color alone,
the star might be an M9 dwarf 
(\cite{Hawley2002},\cite{West2005}) 
with an absolute magnitude $M_i \simeq 15.5 \pm 0.5$.
However, in that case, the distance to the object
would be $\sim 600 \pm 120$ pc, 
and the tangential velocity $v_t \sim 520 \pm 180$ km/s.
The $(i' - z)$ color of an L2 dwarf is not 
very different from our value, and the absolute
magnitude of such a star, $M_i \simeq 17.0 \pm 0.5$,
would yield a lower velocity, 
$v_t \sim 260 \pm 90$ 
km/s,
consistent with the halo population.
The residuals from a simple linear fit to the object's
motion show no sign of parallax,
which indicates that it must be at least
$\sim 20$ pc away from the Sun.

The proper motions computed in this paper
complement those we described in 
\citet{Richmond2009},
since the SXDS lies in the southern galactic
hemisphere and the SDF in the northern.
Together, they comprise a set of 167 stars 
with reliable motions and multicolor photometry.
We have characterized the selection effects
of both samples as a function of magnitude,
proper motion, and (for the SXDS) color.
Our next step will be to compare rigorously 
the properties
of the stars in our samples 
with the predictions made by current models
of galactic populations,
paying special attention to the objects likely 
to be white dwarfs.

% Acknowledgements
\medskip
We thank the staff at the Subaru Telescope 
for their assistance with the observations used in this project.
MWR gratefully acknowledges grant S-03031 from the
JSPS Invitation Fellowship for Research in Japan.
TM is financially supported by the Japan Society for the 
Promotion of Science (JSPS) through the JSPS Research Fellowship. 
Data analysis
was in part carried out on the common use data analysis
computer system at the Astronomy Data Center, ADC,
of the National Astronomical Observatory of Japan.


\begin{thebibliography}{}

\bibitem[Droege et al.\ (2006)]{Droege2006}
   Droege, T. F. et al. 2006, \pasp, 118, 1666
\bibitem[Furusawa et al.\ (2008)]{Furusawa2008}
   Furusawa, H. et al. 2008, \apjs, 176, 1
\bibitem[Hawley et al.\ (2002)]{Hawley2002}
   Hawley, S. L. et al. 2002, \aj, 123, 3409
\bibitem[Kashikawa et al.\ (2004)]{Kashikawa2004}
   Kashikawa, N. et al. 2004, PASJ, 56, 1011
\bibitem[Miyazaki et al.\ (2002)]{Miyazaki2002}
   Miyazaki, S. et al.\ 2002, \pasj, 54, 833
\bibitem[Morokuma et al.\ (2008)]{Morokuma2008}
   Morokuma, T. et al.\ 2008, \apj, 676, 163
\bibitem[Robin et al.\ (2003)]{Robin2003}
   Robin, A. C., Reyl\'{e}, C., Derri\`{e}re, S., \& Picaud, S. 2003, 
        A\&A, 409, 523
\bibitem[Richmond et al.\ (2009)]{Richmond2009}
   Richmond, M. W. et al. 2009, \pasj, 61, 1
\bibitem[Sekiguchi et al.\ (2005)]{Sekiguchi2005}
   Sekiguchi, K. et al. 2005, in 
   Multiwavelength mapping of galaxy formation and evolution
   (Proceedings of the ESO Workshop held at Venice, Italy),
   ed. A. Renzini \& R. Bender (Berlin: Springer-Verlag), 82
\bibitem[Treffers \& Richmond (1989)]{Treffers1989}
   Treffers, R. R. \& Richmond, M. W. 1989, PASP, 101, 725
\bibitem[West et al.\ (2005)]{West2005}
   West, A. A., Walkowicz, L. M., \& Hawley, S. L. 2002, \pasp, 117, 706

\end{thebibliography}
\end{document}